\documentclass[twoside,twocolumn,9pt]{article}
\usepackage{extsizes}
\usepackage[super,sort&compress,comma]{natbib}
\usepackage[version=3]{mhchem}
\usepackage[left=1.5cm, right=1.5cm, top=1.785cm, bottom=2.0cm]{geometry}
\usepackage{balance}
\usepackage{times,mathptmx}
\usepackage{sectsty}
\usepackage{graphicx}
\usepackage{lastpage}
\usepackage[format=plain,justification=justified,singlelinecheck=false,font={stretch=1.125,small,sf},labelfont=bf,labelsep=space]{caption}
\usepackage{float}
\usepackage{fancyhdr}
\usepackage{fnpos}
\usepackage[english]{babel}
\usepackage{array}
\usepackage{charter}
\usepackage[T1]{fontenc}
\usepackage[usenames,dvipsnames]{xcolor}
\usepackage{setspace}
\usepackage[compact]{titlesec}
\usepackage{hyperref}

\usepackage{epstopdf}

\definecolor{cream}{RGB}{222,217,201}

\begin{document}

\pagestyle{fancy}
\thispagestyle{plain}
\fancypagestyle{plain}{

\fancyhead[C]{\includegraphics[width=18.5cm]{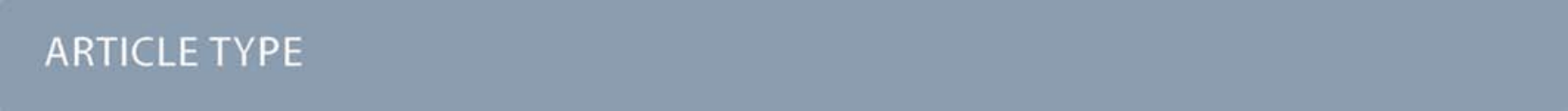}}
\fancyhead[L]{\hspace{0cm}\vspace{1.5cm}\includegraphics[height=30pt]{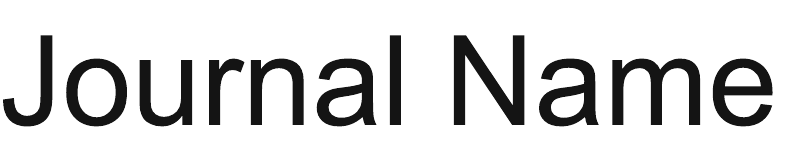}}
\fancyhead[R]{\hspace{0cm}\vspace{1.7cm}\includegraphics[height=55pt]{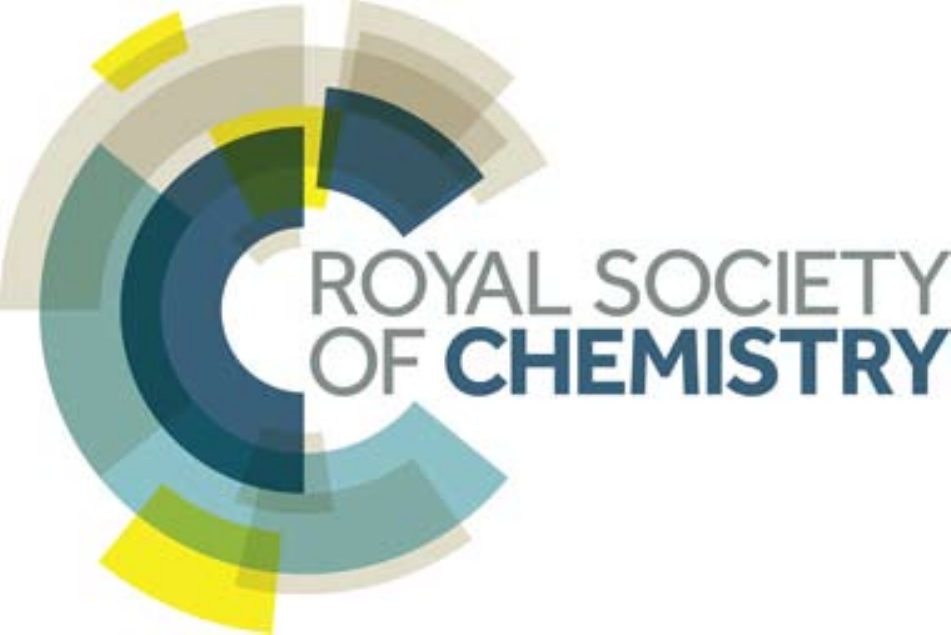}}
\renewcommand{\headrulewidth}{0pt}
 }

\makeFNbottom \makeatletter
\renewcommand\LARGE{\@setfontsize\LARGE{15pt}{17}}
\renewcommand\Large{\@setfontsize\Large{12pt}{14}}
\renewcommand\large{\@setfontsize\large{10pt}{12}}
\renewcommand\footnotesize{\@setfontsize\footnotesize{7pt}{10}}
\renewcommand\scriptsize{\@setfontsize\scriptsize{7pt}{7}}
\makeatother

\renewcommand{\thefootnote}{\fnsymbol{footnote}}
\renewcommand\footnoterule{\vspace*{1pt}%
\color{cream}\hrule width 3.5in height 0.4pt \color{black} \vspace*{5pt}}
\setcounter{secnumdepth}{5}

\makeatletter
\renewcommand\@biblabel[1]{#1}
\renewcommand\@makefntext[1]%
{\noindent\makebox[0pt][r]{\@thefnmark\,}#1}
\makeatother
\renewcommand{\figurename}{\small{Fig.}~}
\sectionfont{\sffamily\Large}
\subsectionfont{\normalsize}
\subsubsectionfont{\bf}
\setstretch{1.125} 
\setlength{\skip\footins}{0.8cm}
\setlength{\footnotesep}{0.25cm}
\setlength{\jot}{10pt}
\titlespacing*{\section}{0pt}{4pt}{4pt}
\titlespacing*{\subsection}{0pt}{15pt}{1pt}

\fancyfoot{}
\fancyfoot[LO,RE]{\vspace{-7.1pt}\includegraphics[height=9pt]{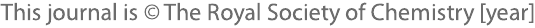}}
\fancyfoot[CO]{\vspace{-7.1pt}\hspace{13.2cm}\includegraphics{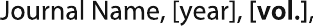}}
\fancyfoot[CE]{\vspace{-7.2pt}\hspace{-14.2cm}\includegraphics{head_foot/RF}}
\fancyfoot[RO]{\footnotesize{\sffamily{1--\pageref{LastPage} ~\textbar  \hspace{2pt}\thepage}}}
\fancyfoot[LE]{\footnotesize{\sffamily{\thepage~\textbar\hspace{3.45cm} 1--\pageref{LastPage}}}}
\fancyhead{}
\renewcommand{\headrulewidth}{0pt}
\renewcommand{\footrulewidth}{0pt}
\setlength{\arrayrulewidth}{1pt}
\setlength{\columnsep}{6.5mm}
\setlength\bibsep{1pt}

\makeatletter
\newlength{\figrulesep}
\setlength{\figrulesep}{0.5\textfloatsep}

\newcommand{\topfigrule}{\vspace*{-1pt}%
\noindent{\color{cream}\rule[-\figrulesep]{\columnwidth}{1.5pt}} }

\newcommand{\botfigrule}{\vspace*{-2pt}%
\noindent{\color{cream}\rule[\figrulesep]{\columnwidth}{1.5pt}} }

\newcommand{\dblfigrule}{\vspace*{-1pt}%
\noindent{\color{cream}\rule[-\figrulesep]{\textwidth}{1.5pt}} }

\makeatother

\twocolumn[
  \begin{@twocolumnfalse}
\vspace{3cm} \sffamily
\begin{tabular}{m{4.5cm} p{13.5cm} }

\includegraphics{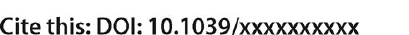} & \noindent\LARGE{Microwave coherent control of ultracold ground-state molecules formed by short-range photoassociation} \\
 & \vspace{0.3cm} \\
 & \noindent\large{Zhonghua Ji,$^{\ast}$\textit{$^{a,b}$} Ting Gong,\textit{$^{a,b}$} Yonglin He,\textit{$^{c}$} Jeremy M. Hutson,\textit{$^{d}$} Yanting Zhao,$^{\dag}$\textit{$^{a,b}$} Liantuan Xiao,\textit{$^{a,b}$} and Suotang Jia\textit{$^{a,b}$}} \\ & \vspace{-0.3cm} \\
\includegraphics{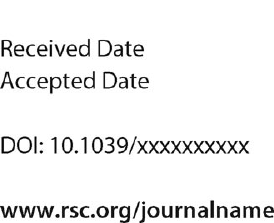} &\noindent\sffamily{We report the observation of microwave
coherent control of rotational states of ultracold $^{85}$Rb$^{133}$Cs molecules formed in their
vibronic ground state by short-range photoassociation. Molecules are formed in the single
rotational state $X(v=0,J=1)$ by exciting pairs of atoms to the short-range state
$(2)^{3}\Pi_{0^{-}} (v=11, J=0)$, followed by spontaneous decay. We use depletion spectroscopy to
record the dynamic evolution of the population distribution and observe clear Rabi oscillations
while irradiating on a microwave transition between coupled neighbouring rotational levels. A
density-matrix formalism that accounts for longitudinal and transverse decay times reproduces both
the dynamic evolution during the coherent process and the equilibrium population. The coherent
control reported here is valuable both for investigating coherent quantum effects and for
applications of cold polar molecules produced by continuous short-range photoassociation.}\\
&\vspace{0.3cm}
 \\
\end{tabular}

 \end{@twocolumnfalse} \vspace{0.6cm}

  ]

\renewcommand*\rmdefault{bch}\normalfont\upshape
\rmfamily
\section*{}
\vspace{-1cm}


\footnotetext{\textit{$^{a}$Shanxi University, State Key Laboratory of Quantum Optics and Quantum Optics Devices, Institute of Laser Spectroscopy, Wucheng Rd. 92,  030006 Taiyuan, China. }}
\footnotetext{\textit{$^{b}$Shanxi University, Collaborative Innovation Center of Extreme Optics, Wucheng Rd. 92, 030006 Taiyuan, China.}}
\footnotetext{\textit{$^{c}$Institute of Theoretical Physics, School of Physics and Electromechanical Engineering, Hexi University, Zhangye 734000, P.R. China. }}
\footnotetext{\textit{$^{d}$Joint Quantum Centre (JQC) Durham-Newcastle, Department of Chemistry, Durham University, South Road, Durham DH1 3LE, United Kingdom \\E-mail: jzh@sxu.edu.cn($^{\ast}$), zhaoyt@sxu.edu.cn($^{\dag}$)}}

\rmfamily 

\section*{Introduction}
Recent decades have witnessed fast developments in the study of ultracold  molecules, which are of
great interest in both physics and chemistry
\cite{Dulieu-pccp,Chem.Rev.2012-Control,Chem.Rev2012-PA,Moses-2017-NP,Bohn-2017-science}. Ultracold
polar molecules have abundant internal states and interact via strong, anisotropic, and long-ranged
dipolar interactions. They have potential applications in ultracold chemistry \cite{Julienne-pccp,
Krems-pccp-2008,WangdDJ2018ScienceAdv,Cornish2019NC,Ni2019Science}, precision
measurement\cite{Chin_2009,Safronova2018RMP}, quantum simulation \cite{Blackmore_2018} and quantum
computation \cite{Demille2002PRL,kk2018}.

All these applications require cold polar molecules in a well-defined initial state. Such molecules
may be produced in a variety of ways. There have been rapid recent developments in direct laser
cooling and magneto-optical trapping (MOT) to produce ultracold polar molecules in a single quantum
state \cite{DeMille-2014-nature, Tarbutt-2017-np, Doyle-2017-prl}. However, this technique is
limited to a small class of molecules with nearly closed laser-cooling transitions, and currently
produces only very low phase-space densities. An alternative approach is to form molecules from
pairs of ultracold atoms by magnetoassociation using a magnetic field ramp. Such molecules are
initially in weakly bound Feshbach states of the electronic ground state, but it is often possible
to transfer them coherently to the vibronic ground state by stimulated Raman adiabatic passage
(STIRAP) \cite{Bergmann-1998-rmp}. In favourable cases, this can produce substantial densities of
molecules in a single hyperfine and Zeeman state \cite{Ni-2008-science, Takekoshi-2014-prl,
Molony-2014-prl, Park-2015-prl, Guo-2016-prl, Timur-2017-prl}. A further class of methods is based
on photoassociation (PA) using laser beams. One-photon PA associates pairs of atoms to form cold
molecules in an excited electronic state, usually highly vibrationally excited. These molecules
soon decay spontaneously to the ground excited state; they are usually still in highly excited
vibrational states, but may be transferred to the vibronic ground state either by a pump-dump
scheme \cite{Memille2005PRL} or by STIRAP \cite{Aikawa-2010-prl}.

Coherent methods can form molecules only once per experimental cycle. By contrast, methods that
involve spontaneous emission allow accumulation of molecular density. In particular, short-range
photoassociation offers a simple optical pathway from atoms to molecules; it allows accumulation of
cold molecules in the vibronic ground state, and can be applied to systems that are not amenable to
magnetoassociation. Short-range PA has been implemented for various heteronuclear molecules
\cite{PhysRevLett.101.133004, PhysRevA.84.061401, PhysRevA.86.053428, Demille-2015-pra,
PhysRevA.94.062510}. We have recently formed polar $^{85}$RbCs molecules in this way, trapped them
optically, and measured atom-molecule collision rates \cite{LiZH2018pccp}, while Passagem \emph{et
al.} have developed a special laser that can simultaneously photoassociate $^{85}$Rb atoms and
trapping the resulting Rb$_{2}$ homonuclear molecules \cite{Marcassa2019PRL}.

\begin{figure*}[t]
\centering
\includegraphics[width=0.76\textwidth]{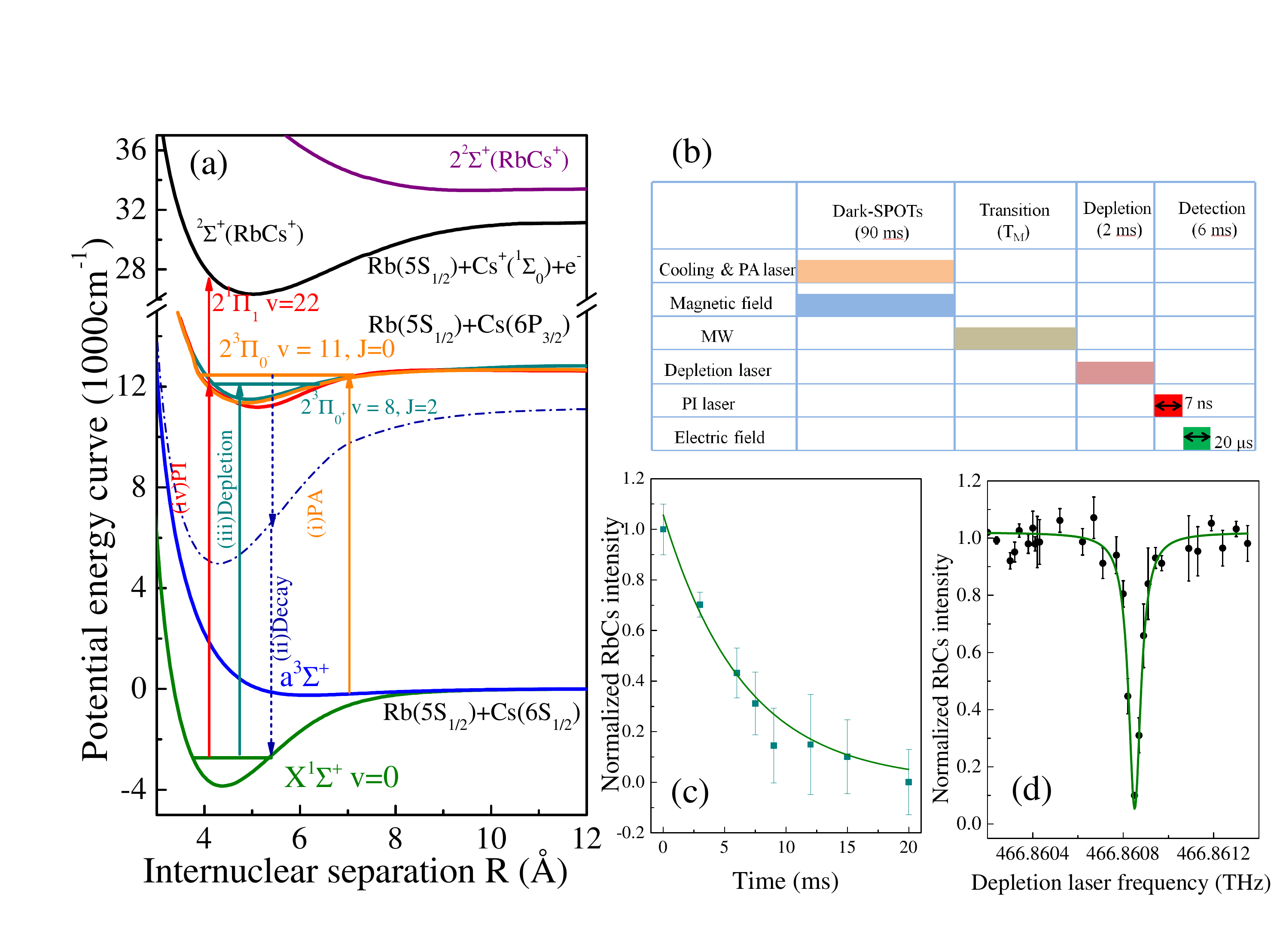}
\caption{(Color online) Experimental overview. (a) Optical pathways to produce and detect ultracold
$^{85}$Rb$^{133}$Cs molecules in the state $X^{1}\Sigma^{+}(v=0)$. (b) Time sequence. (c) Lifetime measurement of formed molecules in the absence of both the depletion laser and the MW field. (d) Depletion spectroscopy measurement in the absence of MW field. }
	\label{fig1}
\end{figure*}

Once molecules have been produced in a single quantum state, coherent control is needed. Polar
molecules have allowed microwave (MW) transitions between neighbouring rotational states
\cite{Aldegunde-2009-pra}, which may be driven with very high resolution. Coherent control using
such transitions has been achieved both for molecules produced by STIRAP, including
$^{40}$K$^{87}$Rb \cite{PhysRevLett.104.030402}, $^{23}$Na$^{40}$K \cite{PhysRevLett.116.225306,
Park-2017-science}, $^{87}$Rb$^{133}$Cs \cite{PhysRevA.94.041403}, $^{23}$Na$^{87}$Rb
\cite{PhysRevA.97.020501}, and for CaF produced by direct laser cooling
\cite{PhysRevLett.120.163201}. Such control is at the heart of nearly all proposals for
applications, such as simulating quantum magnetism \cite{Gorshkov-2011-prl, Ryan-2006-prl},
coupling quantum qubits \cite{Demille-2002-prl, Andre-2006-np}, controlling state-dependent
chemical reactions \cite{Ospelkaus-2010-science}, inducing dipolar interaction for topological
phase \cite{Cooper-2009-prl}, enhancing evaporative cooling \cite{Avdeenkov-2012-pra} and synthetic
dimensions \cite{Sundar-2018-sr}.

Here we report coherent control of rotational states of ultracold polar molecules produced in the
lowest vibronic state by continuous photoassociation. We observe clear Rabi oscillations between
neighbouring rotational states. We use a density-matrix formalism that accounts for longitudinal
and transverse decay times to analyze the evolution of the population distributions, determine the
coherence time, and understand the equilibrium state.

\section*{Experimental setup}

A full description of our apparatus has been given previously \cite{Li-2018-oe}.  We
therefore give only a summary of the main procedures and parameters here; further details can be
found in ref.\ \cite{Li-2018-oe}.

The precooled atom samples are prepared as before, but here we photoassociate via an excited
molecular state that decays to a particularly simple rotational distribution in the lowest vibronic
state. This facilitates subsequent control of the quantum state. We also use a
different time sequence, which is shown in Fig.\ \ref{fig1}(b). This separates the
microwave coupling  step from population measurement by depletion spectroscopy,
allowing us to investigate the coherence of the microwave transition.

Under a vacuum background pressure around 3$\times$$10^{-7}$ Pa and at a magnetic gradient around
15 G/cm, we trap a mixed atomic cloud that consists of 1$\times$$10^{7}$ Rb atoms in the state
$5S_{1/2}\ (F = 2)$ and  2$\times$$10^{7}$ Cs atoms in the state $6S_{1/2}\ (F = 3)$. The number
densities of Rb and Cs atoms are 8$\times$$10^{10}$ cm$^{-3}$ and 1$\times$$10^{11}$ cm$^{-3}$,
respectively. The translational temperature of the mixture is measured by time-of-flight imaging to
be around 100 $\mu$K.

We carry out photoassociation using the optical pathways shown in Fig.\ \ref{fig1}(a), with
potential energy curves based on the results of Refs.\ \cite{Fahs-2002-jpb} and
\cite{Allouche-2000-jpb}. The chosen intermediate molecular state is $2^{3}\Pi_{0^{-}}$ ($v=$11,
$J=$0). As shown by Shimasaki \emph{et al.} \cite{Demille-2015-pra}, this state decays to
$X^{1}\Sigma^{+}(v=0)$ by two-photon cascade. The molecules formed are then ionized
by a tunable pulsed dye laser, accelerated by a pulsed electric field, and finally detected by a
pair of micro-channel plates. In the absence of both the depletion laser and the MW field, the
lifetime is measured to be 6.6(6) ms, as shown in Fig.\ \ref{fig1}(c). When parity is conserved,
the two-photon decay produces only one rotational state, $J=1^-$, although Ref.\
\cite{Demille-2015-pra} observed small populations in $J=0^+$ and $2^+$ as well, due to Stark
mixing induced by the residual static electric field in their experiment.

Figure \ref{fig1} (d) shows the result of depletion spectroscopy for the
molecules we produce in the absence of the MW field. The interaction time and
intensity of the depletion laser are 2~ms and 1~mW/mm$^{2}$ respectively.
The single transition observed may be assigned as $X^{1}\Sigma^{+}(v=0, J=1)$ to
$2^{3}\Pi_{0^{+}}$ ($v=8, J=2)$, based on the assignments in Fig.\ 2(b) of Ref.
\cite{Li-2018-oe}.  The fractional depletion is greater than 90\% and the transition
from $J=3$ to 4 is not observed. This demonstrates that only the $J=$1 state of $X^{1}\Sigma^{+}\
(v=0)$ is  significantly populated in the present experiment. Transitions originating
in $J=0$ and 2 are also not observed, confirming that the influence of Stark mixing is not
important here. The full width at half maximum (FWHM),
obtained by fitting to a Lorentz lineshape, is found to be 73(5) MHz and is attributed to power
broadening due to the depletion laser.

\section*{Experimental results and analysis}
\begin{figure}[t]
\centering
\includegraphics[width=0.45\textwidth]{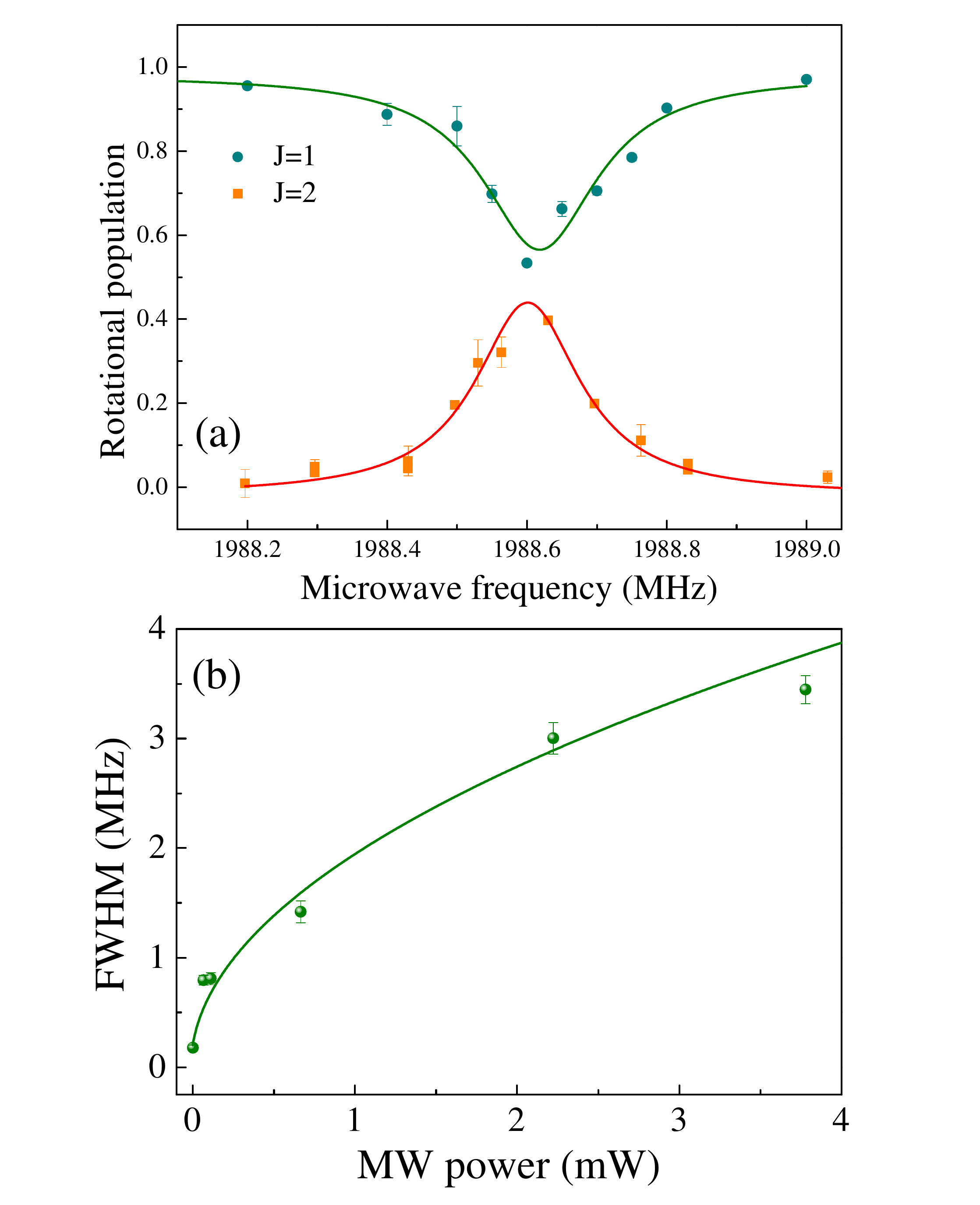}
\caption{(Color online) (a) The observed microwave transition in ultracold RbCs molecules,
monitored with the depletion laser locked to the transition $2^{3}\Pi_{0^{+}}$ ($v=8,
J=2)$$\leftarrow$$X^{1}\Sigma^{+}(v=0, J=1)$ (green squares) and the transition
$2^{3}\Pi_{0^{+}}$ ($v=8, J=3)$$\leftarrow$$X^{1}\Sigma^{+}(v=0, J=2)$ (orange squares).
The curves are fitted to a Lorentzian lineshape. (b) The FWHM width $\gamma$ as a function of
microwave power.}
	\label{fig2}
\end{figure}

After production of molecules in the state $X^{1}\Sigma^{+}(v=0, J=1)$, we irradiate them with
microwave radiation close to the $J=1\rightarrow 2$ transition. The MW field is
 produced with a homemade coil and has no well-determined quantization axis. The
resulting rotational population distributions are shown as a function of MW frequency in Fig.\
\ref{fig2}(a). The interaction time is chosen to be 2~ms, which is longer than the coherence time.
The microwave intensity is around 1~$\mu$W, which is enough lower than the saturation power and can simultaneously
ensure sufficient signal-to-noise. We measure the radiant power using a microwave power meter
(NRP-Z51, R$\&$S) with a circle probe of diameter 2.3~cm. Since the power at the atomic cloud
cannot be measured directly, we use the measured value at an equivalent distance from the homemade
radiant coil. To measure the population in $J=$1, the frequency of the depletion laser shown in
Fig.\ \ref{fig1} is locked at the transition between $X^{1}\Sigma^{+}(v=0, J=1)$ and
$2^{3}\Pi_{0^{+}}(v=8, J=2)$, while for $J=$2 it is locked at the corresponding transition between
$J=$2 and $J=$3. The population is obtained from one minus the the ratio of the
intensity of RbCs ions in the presence of the depletion laser to that in its absence. Fitting to Lorentzian lineshapes gives a resonant microwave frequency $\nu_{12}=1988.62(1)$
MHz and  FWHM $\gamma=0.20(4)$ MHz from the
population in $J=1$ and $\nu_{12}=1988.60(1)$ MHz and $\gamma=0.18(1)$ MHz from that in $J=2$. We
can use either rotational state to probe coherent control. In the following, we focus on the
population of the state $X^{1}\Sigma^{+}(v=0, J=1)$. Figure \ref{fig2}(b) shows the value of
$\gamma$ from the population in $J=$1 as a function of MW power. We use a simple model
$\gamma=\gamma_{0}\sqrt{1+P/P_\textrm{sat}}$ \cite{Chen-2017-pra} to fit the
experimental data, yielding $\gamma_{0}$=0.20(8) MHz and $P_\textrm{sat}$=0.011(1) mW.

\begin{figure}[!t]
\centering
\includegraphics[width=0.45\textwidth]{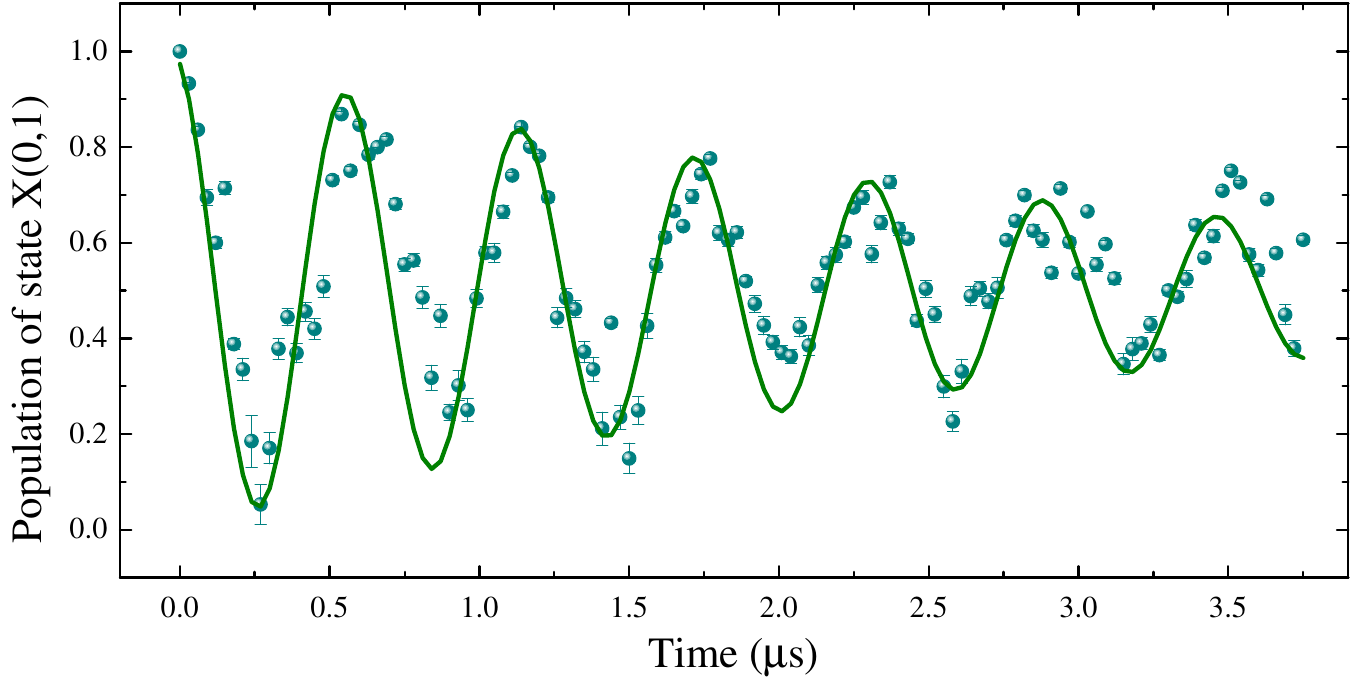}
\caption{(Color online) Evolution of the population of the initial state of the molecules, $J=1$.
The curve is a simulation based on Eq. (1) and (2). Each  point represents the mean of 36
measurements. }
	\label{fig3}
\end{figure}

Figure \ref{fig3} shows the population in $J=1$ as a function of MW irradiation time. The MW
frequency is fixed at the central value fitted in Fig.\ \ref{fig2}(a). The measured MW power is
10~mW. The measured population shows a clear Rabi oscillation. We treat the two-level system
theoretically using a density-matrix formalism under the electric-dipole and rotating-wave
approximations. The time evolution is written as a pair of coupled equations \cite{Rand2010Book},
\begin{eqnarray}
\mathop {{\rho _{21}}}\limits^ \cdot   =  - ({\Gamma _2} + i\Delta ){\rho _{21}} - i\frac{\Omega }{2}({\rho_{22}} - {\rho _{11}}); \\
\mathop {{\rho _{22}}}\limits^ \cdot   = {\Gamma _1}(\rho _{22}^0 - {\rho _{22}}) + i\frac{\Omega }{2}{\rho _{12}} - i\frac{\Omega }{2}{\rho _{21}},
\label{eq01}
\end{eqnarray}
with parameters ${\Gamma _1} = 1/T_{1}$ and ${\Gamma _2} = 1/T_{2}$, where $T_{1}$ and $T_{2}$ are
the longitudinal and transverse decay times. These times characterise the timescales for changes in
population and for decoherence, respectively. $\Delta$ is the detuning of the applied fields and
$\Omega=\mu_{12} E/\hbar$ is the Rabi frequency, where $\mu_{12}$ is the transition dipole moment
(TDM) and $E$ is the MW amplitude. The coupled differential equations are solved
numerically, with the initial condition $\rho_{11}^0=1$ at $t=0$ and the constraints
${\rho_{11}}+{\rho_{22}}=1$ and ${\rho_{12}}=\rho_{21}^*$.

The curve in Fig.\ \ref{fig3} shows the numerical simulation of $\rho _{11}$. In the simulation,
the MW frequency is on resonance and the longitudinal decay time $T_{1}$ is chosen to be 6.6~ms,
which is the measured lifetime of the molecules from Fig.\ \ref{fig1}(c).  A satisfactory simulation is obtained with Rabi frequency $\Omega= 1.70$ MHz and coherence time $T_2 = 1.5$ $\mu$s. This
value of ${T_2}$ is more than two orders of magnitude lower than in other systems
\cite{PhysRevLett.104.030402, PhysRevLett.116.225306, Park-2017-science, PhysRevA.94.041403,
PhysRevA.97.020501, PhysRevLett.120.163201}, but is still sufficient to allow several coherent
manipulations and to investigate coherence effects. The factors limiting coherence arise mainly
from the higher temperature of the precooled atoms in the present work, from inelastic molecular collisions \cite{LiZH2018pccp} and
from unresolved hyperfine structure.

The nuclei $^{85}$Rb and $^{133}$Cs have spins $I_\textrm{Rb}=5/2$ and
$I_\textrm{Cs}=7/2$. The dominant hyperfine interactions in states with $J>0$ are nuclear electric
quadrupole interactions, with coupling constants $(eQq)_\textrm{Rb}=-1.661$ MHz and
$(eQq)_\textrm{Cs}=0.054$ MHz, obtained by Aldegunde and Hutson \cite{Aldegunde2017PRA} from
electronic structure calculations using density functional theory. There are several other
hyperfine terms, but their effect is at the kHz level. The allowed microwave lines for
$J=1\rightarrow2$ are expected to be spread over about 750 kHz; this is consistent with the
observed coherence time  ${T_2}=1.5\ \mu$s and may be compared with the observed FWHM
$\gamma = 0.200(80)$~MHz.

\begin{figure}[!htb]
\centering
\includegraphics[width=0.45\textwidth]{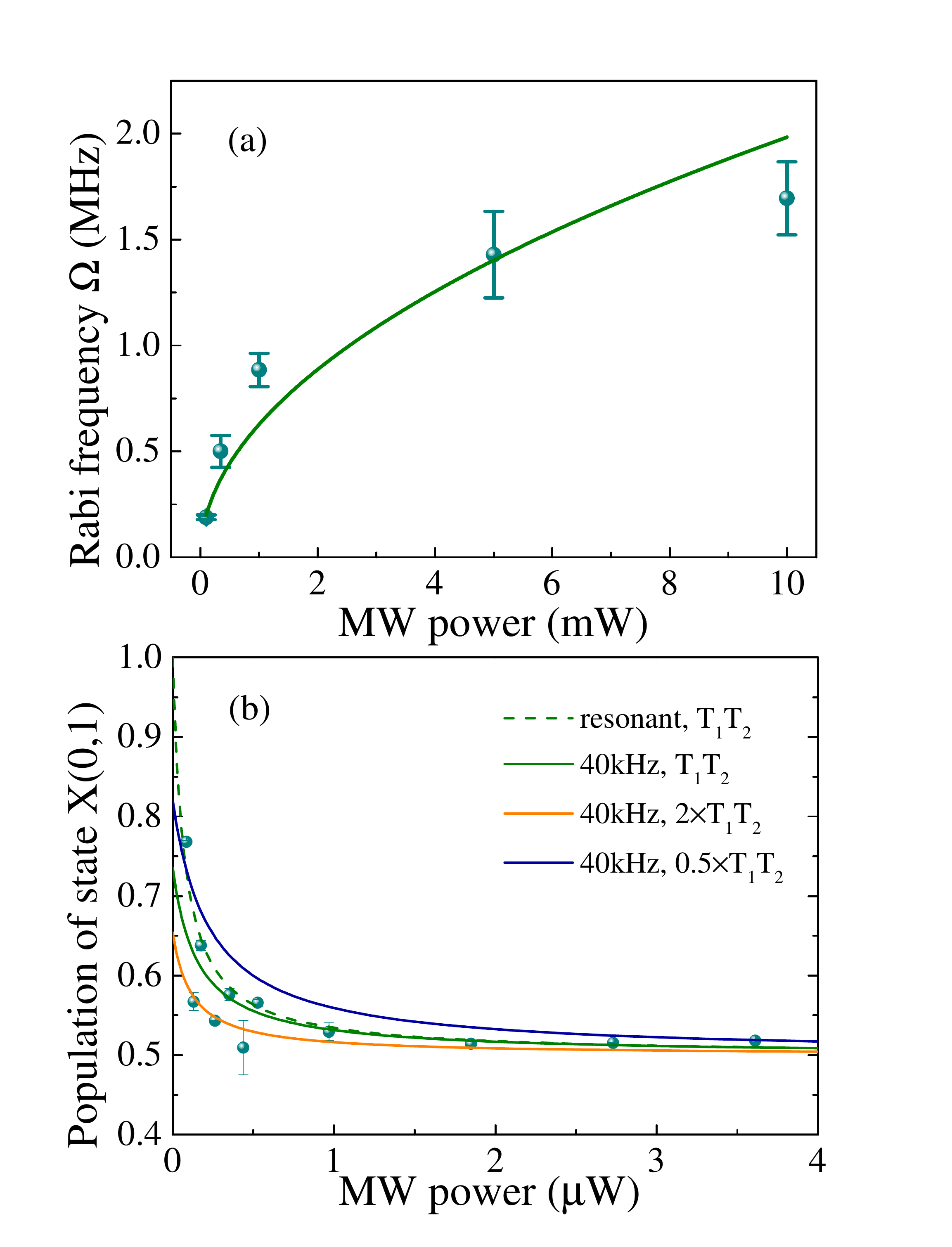}
\caption{(Color online)  (a) Dependence of the Rabi frequency extracted from the coherence measurements on MW power.
(b) The population of RbCs molecules in the initial state as a function of MW power, after irradiation for 2~ms.}
	\label{fig4}
\end{figure}

The coherence time ${T_2}$ is independent of the experimental conditions, but the Rabi frequency
$\Omega$ can be controlled. It depends externally on the MW amplitude and internally  on the TDM.
The MW amplitude $E$ and power $P$ are related to the intensity $I$ by $I = \frac{1}{2}c n
\epsilon_{0}  E^{2} = P/\pi r^{2}$, so that $\Omega=(2\mu_{12}^{2}P/c n \epsilon_{0} \pi r^{2}
\hbar^{2})^{1/2}$. The Rabi frequency obtained from coherence measurements is shown as a function
of MW power in Fig.\ \ref{fig4}(a). The fitted curve gives $\mu_{12}=0.53(9)$ Debye. Here the error
0.09 Debye includes only the statistical error from the fit. There is an additional uncertainty in
the TDM due to uncertainty in the MW power at the position of the atomic cloud, which
we estimate as $\pm0.03$ Debye.

Our experiments use unpolarized
microwaves, but in an isotropic environment the intensities may be obtained by considering
plane-polarized light with the electric vector along a single direction \cite{Hougen1970Book}. For
$z$ polarization, the TDM between the $M=0$ components of $J=1$ and $2$  is
$\mu_{12} =(4/15)^{1/2} \mu(v,J)$,\cite{Brink-Satchler1994Book} where $\mu(v,J)$ is the
molecule-fixed dipole moment. This has been measured for $^{87}$RbCs in  its
rovibronic ground state  as $\mu(0,0)=1.225(11)$ Debye
\cite{Molony-2014-prl}. To obtain the corresponding value for $^{85}$RbCs, we solve the vibrational
Schr\"odinger equation for each isotopolog using the ground-state RbCs interaction potential of
Takekoshi \emph{et al.} \cite{PhysRevA.85.032506} and evaluate the expectation values
$\mu(v,J)$ using the dipole-moment function of Fedorov \emph{et al.}
\cite{JCP.140.184315} The value obtained for  $\mu(0,0)$ in $^{85}$RbCs is only about
9 parts in $10^7$ smaller than for $^{87}$RbCs. The absolute value $\mu (0,0)=
1.215$ D is less accurate than experiment, but the ratio between isotopologs is reliable. The
dependence on rotational state is also negligible; it may be approximated by
$\mu(v,J) = \mu(v,0) + D_{\mu v} J(J+1)$, where $D_{\mu 0} = 2.34 \times 10^{-7}$ Debye for
$^{85}$RbCs. The experimental value of $\mu(0,0)$ \cite{Molony-2014-prl} corresponds to
$\mu_{12}=0.633$ Debye, which is consistent with the present result in view of the uncertainty in
the measured MW power.

At times much longer than the coherence time, the system reaches equilibrium and the population of
the initial state $X^{1}\Sigma^{+}(v=0, J=1)$ becomes stable. Figure \ref{fig4}(b) shows the
measured population of the state $X^{1}\Sigma^{+}(v=0, J=1)$ as a function of MW power for an
irradiation time of 2~ms, which is long enough for equilibrium to be established. It may be seen
that the steady-state value is a little larger than 0.5. Ref.\ \cite{Sanz-2006-jcp} gives the
steady-state population for the ideal resonant frequency. However, in a real experiment the
detuning $\Delta$ is finite, though small, so here we use the generalized Rabi frequency $\tilde
{\Omega}=\sqrt{\Omega^{2}+\Delta^2}$ in place of the Rabi frequency,
\begin{equation}
\rho^\textrm{eq}_{11}=\frac{1}{2}\left[1+\frac{\tilde{R}_{3}}{(1+T_{1}T_{2}(\Omega^{2}+\Delta^2))}\right].
\label{eq1}
\end{equation}
Here $\tilde{R}_{3}=(1-e^{-\hbar \omega_{0}/k_\textrm{B}T})/(1+e^{-\hbar
\omega_{0}/k_\textrm{B}T})$ indicates the degree of mixedness of the reduced density matrix at
temperature $T$ in the absence of the external MW field and $\omega_{0}=(E_{2}-E_{1})/\hbar$ is the
resonant angular frequency. At the temperature of our experiment, $T=100\ \mu$K, $\tilde{R}_{3}$ is
approximated as 1. The green dashed line and solid line in Fig.\ \ref{fig4}(b) show the simulated
results when the detuning is zero (\emph{i.e.}\ resonant) and 40~kHz (the maximum uncertainty in
the FWHM of the MW spectra), respectively, using the lifetime $T_{1}=7$~ms and coherence
time $T_{2}=4\ \mu$s. Since there are large uncertainties in $T_{1}$ and $T_{2}$ and the product of
them influences the steady-state population from Eq. \ref{eq1}, we have repeated the simulation for
$\Delta$=40~kHz with the experimental value of $T_{1}T_{2}$  halved and doubled. Figure
\ref{fig4}(b) shows that nearly all the measured populations are within the range of the simulated
curves, which supports the theoretical model and estimates of uncertainty.

\section*{Conclusion}

In conclusion, we have demonstrated MW coherent control of ultracold polar $^{85}$RbCs molecules
formed by continuous short-range photoassociation from a cold atomic mixture. We observe clear Rabi
oscillations and simulate them by adding decay terms to the classical Hamiltonian of a two-level
system in a monochromatic electric field. The transition dipole moment measured between adjacent
rotational states is consistent with the theoretical value. The coherence time and lifetime of the
ground state molecules are limited by the relatively high temperature and the fact that the
molecules are in an unpolarized state with unresolved hyperfine structure.
Nevertheless, the coherence time is long enough to investigate both the dynamic evolution during
the coherent process and the equilibrium population. Techniques such as Raman sideband cooling are
expected to improve the coherence properties by allowing preparation of the atomic sample at lower
temperature and in a polarized state.

\section*{Conclusion}
There are no conflicts to declare.

\section*{Acknowledgments}
This work was supported by National Key R$\&$D Program of China (Grant No. 2017YFA0304203), Natural
Science Foundation of China (Nos. 61675120, 61875110), NSFC Project for Excellent Research Team
(No. 61121064), Shanxi $``$1331 Project$"$ Key Subjects Construction, PCSIRT (No. IRT$_{-}$17R70),
111 project (Grant No. D18001) from China. Y. L. He is supported by NSF of China (Grant No.
11464010) and J. M. Hutson is supported by the U.K. Engineering and Physical Sciences Research
Council (EPSRC) Grants No.\ EP/N007085/1, EP/P008275/1 and EP/P01058X/1.


\scriptsize{
\bibliography{Refs-4.1.1} 
\bibliographystyle{rsc} } 

\end{document}